\documentclass[aps,prd,amssymb,cite,%,showkeys,showpacs
amsfonts,epsf,preprintnumbers,nofootinbib,superscriptaddress]{revtex4}
\usepackage{amsmath,amssymb,amsfonts}
\usepackage{bm}
\usepackage[dvips]{graphicx}
\usepackage{bm,latexsym,amsmath,amssymb,amsfonts}
\usepackage[usenames,dvipsnames]{color}
\usepackage[colorlinks=true,linkcolor=blue]{hyperref}
\usepackage{color}
\usepackage{soul}
\soulregister\cite7
\soulregister\ref7
\soulregister\pageref7
\usepackage{epsfig}
\usepackage{braket}
\usepackage[mathscr]{eucal}
\usepackage{cancel}
\usepackage{mathrsfs}
\usepackage{pgf, tikz}
\usepackage{slashed}
\usetikzlibrary{arrows,automata}
\usepackage{pgfplots}
\pgfplotsset{compat=newest}

\usepackage{amsthm}

\theoremstyle{remark}

\newcommand{\be}[1]{\begin{equation} \label{#1}}
\newcommand{\ee}{\end{equation}}
\newcommand{\bea}{\begin{eqnarray}}
\newcommand{\bean}{\begin{eqnarray*}}
\newcommand{\eea}{\end{eqnarray}}
\newcommand{\eean}{\end{eqnarray*}}
\newcommand{\ba}{\begin{array}}
\newcommand{\ea}{\end{array}}
\newcommand{\bel}{\begin{align}}
\newcommand{\eel}{\end{align}}

\newcommand{\tcb}{\textcolor{blue}}

\newcommand{\tcg}{\textcolor{green}}

\begin{document}

\title{Electromagnetism from two matter spaces: mutual helicity and the nondegenerate completion}
\author{Hyeong-Chan Kim}
        \email{Corresponding Author, hckim@ut.ac.kr}
	 \affiliation{School of Liberal Arts and Sciences, Korea National University of Transportation, Chungju 27469, Korea}
%\date{today}
\begin{abstract}
We show that generic Maxwell fields can be represented within the
matter-space framework by introducing two independent matter-space flows.
In the one-flow formulation the electromagnetic field strength is the
pull-back of a two-form on a three-dimensional matter space and therefore
satisfies $F\wedge F=0$, so that a single flow captures only a degenerate,
helicity-carrying sector. The minimal completion is obtained by writing
$$F=G^{(1)}+G^{(2)},$$
where each sector field $G^{(I)}$ is the pull-back of a matter-space
two-form from an independent flow. Each sector is individually degenerate,
$G^{(I)}\wedge G^{(I)}=0$, while the full field satisfies $F\wedge
F=2G^{(1)}\wedge G^{(2)}$; the invariant excluded by the one-flow theory is
thus recovered as an inter-flow quantity. We interpret this structure in
terms of mutual helicity: the total helicity decomposes into two
self-helicity contributions and a mutual term whose exterior derivative is
$F\wedge F$. Hence a configuration with vanishing mutual helicity lies in
the degenerate sector, whereas a nondegenerate Maxwell field necessarily
carries a nontrivial mutual helicity. A variational principle for the
total field recovers the sourced Maxwell equation, the two matter-space
variations combining into the full equation whenever the sector kernels
intersect trivially. Generic electromagnetism is thereby reconstructed as
the minimal coupled completion of two individually degenerate matter-space
sectors.
\end{abstract}

\maketitle
\onecolumngrid

%=====================================================
\section{Introduction}
\label{sec:introduction}
%=====================================================

In our previous work~\cite{Ho2026CQG}, electromagnetism was reformulated within the matter-space framework of relativistic fluid dynamics~\cite{AnderssonComer2007,Carter1973,Carter1989} in terms of a single matter-space flow. 
The central result of that construction was that a one-flow matter space naturally generates a constrained electromagnetic sector: the induced field strength is described by a simple or degenerate two-form, and the associated conserved matter-space current is most naturally interpreted as a helicity current. 
In this sense, the one-flow formulation provides a geometrically natural description of an elementary helicity-carrying electromagnetic sector.
Null fields provide an important subclass of simple electromagnetic two-forms~\cite{Robinson1961}.

At the same time, the one-flow theory makes clear that such a sector does not exhaust the full local content of Maxwell theory. 
A generic electromagnetic field in four spacetime dimensions is not simple: in general it has rank four and carries nonvanishing invariant structure beyond that of a single degenerate sector. 
This immediately raises a natural question: if the one-flow construction captures only an elementary helicity sector, what is the minimal matter-space completion required to recover a generic Maxwell field?

The purpose of the present work is to answer this question. 
We show that the natural completion of the one-flow matter-space formulation is a \emph{two-flow} construction. 
The starting point is the elementary but crucial algebraic fact that, in four spacetime dimensions, a generic electromagnetic two-form can be written locally as the sum of two simple two-forms. 
This suggests that a generic Maxwell configuration should be interpreted not as a single matter-space sector, but as a superposition of two one-flow sectors, each carrying its own helicity current and admitting its own matter-space realization. 
%The one-flow theory is thus reinterpreted as the \tcr{elementary building block} of the general electromagnetic field.

However, the passage from one flow to two flows is not merely a matter of doubling degrees of freedom. 
Once the field is decomposed into two simple sectors, the helicity of the total configuration also decomposes into three parts: two sector self-helicities and an additional \emph{mutual helicity} term.
We will show that the mutual term has invariant content: its exterior derivative is the four-form \(F\wedge F\).  
Therefore a mutually helicity-free configuration cannot represent the generic sector with \(F\wedge F\neq0\). 
This is the precise sense in which the nondegenerate part of the Maxwell field is tied to the coupling between the two simple sectors.

This provides a new interpretation of the local structure of the Maxwell field. 
Rather than viewing a generic electromagnetic configuration as a single object with six independent components, the present framework resolves it into two individually degenerate helicity sectors together with the mixed inter-sector structure required for a nondegenerate field.
From this viewpoint, the one-flow matter-space theory is not a failed attempt to recover all of electromagnetism from a single flow; rather, it identifies the elementary constituent sector out of which generic electromagnetism is assembled.

The two-flow perspective also clarifies the role of electromagnetic helicity. 
In the one-flow theory, the preferred conserved matter-space current is naturally aligned with the helicity current of that sector. 
In the multi-flow completion, each simple sector carries its own self-helicity, while the full field acquires additional structure through mutual helicity.
%This suggests that \tcr{the fundamental local organization of the Maxwell field is more naturally described in terms of helicity sectors and their couplings than in terms of a single unconstrained gauge potential alone. }
The present construction therefore does not merely rewrite Maxwell theory in unfamiliar language; it isolates a geometrically meaningful decomposition of the field into elementary matter-space sectors and identifies the precise obstruction to reducing a generic field to a sum of independent one-flow contributions.

The main goal of this paper is therefore not to rederive the one-flow results, but to establish the minimal completion required for generic electromagnetism and to understand its physical content. 
Concretely, we will show that a generic local Maxwell field admits a decomposition into two simple sectors; that each sector may be regarded as a one-flow matter-space sector; and that the mutual helicity between them is necessary for the total field to remain nondegenerate. 
The outcome is a sharpened geometric picture of electromagnetism: the one-flow sector captures the elementary helicity structure, whereas the full Maxwell field is obtained only through the mutual coupling of two such sectors.

This paper is organized as follows. 
In Sec.~\ref{sec:algdecomposition}, we review the local algebraic structure of a generic electromagnetic two-form in four dimensions and formulate its decomposition into two simple sectors. 
Because each matter-space sector must be closed, we show that the relevant decomposition is the one supplied by Darboux's theorem rather than by the metric principal-plane splitting. 
In Sec.~\ref{sec:principalplanesmatterspace}, we introduce the corresponding two-flow matter-space presentation, assigning an independent matter space and helicity current to each simple sector. 
In Sec.~\ref{sec:selfmutualGnotation}, we analyze the decomposition of the total helicity into self and mutual parts and prove that the vanishing of the mutual contribution forces the total field back into a degenerate sector. 
In Sec.~\ref{sec:twoflowaction}, we construct the action for the electromagnetic field and show that the two flow construction naturally reconstruct the electromagnetic source equation.
In Sec.~\ref{sec:discussion}, we discuss the physical meaning of this result and explain how the generic nondegenerate Maxwell field may be understood as the minimal coupled completion of two elementary one-flow helicity sectors.

%==========================================\\
\section{Algebraic decomposition of a general electromagnetic field}
\label{sec:algdecomposition}

The starting point of the multi-flow extension is the elementary algebraic fact that, in four spacetime dimensions, every two-form may be written locally as the sum of at most two simple two-forms. 
Equivalently, given a local electromagnetic field strength $F_{ab}$, there exist simple two-forms $G^{(1)}_{ab}$ and $G^{(2)}_{ab}$ such that
\begin{equation}
F_{ab}=G^{(1)}_{ab}+G^{(2)}_{ab}.
\label{eq:Fsum}
\end{equation}
In four spacetime dimensions such a decomposition always exists locally for
a generic field, and each summand can be chosen to be both simple and
closed; the precise statement and the appropriate normal form are given in
Sec.~\ref{subsec:darboux}.
Each simple closed term may then be interpreted as a one-flow-type
electromagnetic sector.

This immediately motivates the two-flow completion: a generic rank-four electromagnetic field is represented as the sum of two rank-two sectors, each of which is individually of one-flow type. 
In this sense, the one-flow theory describes the elementary building blocks of electromagnetism, while the generic Maxwell field is obtained only after superposing two such sectors.

It is important to distinguish two levels of structure here. 
First, there is the purely algebraic statement that a general two-form is the sum of two simple ones. 
Second, there is the geometric claim that each simple sector admits a matter-space realization. 
In the present framework, we take the latter as the natural extension of the one-flow theory: each simple two-form sector is assigned its own matter space and its own helicity current.

%The algebraic decomposition of a non-null electromagnetic two-form into two simple bivectors is closely related to the standard canonical classification of electromagnetic fields. 
%What is new in the present construction is not this algebraic fact itself, but its interpretation within the matter-space framework: the two simple bivectors are promoted to two one-flow helicity sectors, and \tcr{the invariant $\bm{F}\wedge \bm{F}$ is identified as the gauge-invariant local density associated with their mutual helicity.}

%=====================================
\subsection{Closed simple decomposition and Darboux's theorem}
\label{subsec:darboux}

For the decomposition~\eqref{eq:Fsum} to serve as the starting point of a
matter-space construction, the two summands cannot be arbitrary simple
two-forms; they must satisfy two conditions.  First, each sector must be
simple,
\begin{equation}
G^{(I)}\wedge G^{(I)}=0 ,
\end{equation}
so that it has rank two and can be realized as the pull-back of a two-form
from a three-dimensional matter space.  Second, and decisively, each sector
must be closed,
\begin{equation}
dG^{(I)}=0 ,
\label{eq:sectorclosed}
\end{equation}
because in the matter-space framework the sector two-form is the pull-back
of a closed matter-space two-form and therefore admits a sector potential
$G^{(I)}=dX^{(I)}$, as in Eqs.~\eqref{closed} and~\eqref{G:dX}.  A
decomposition into simple but non-closed summands cannot be lifted to two
independent matter-space flows, since a non-closed $G^{(I)}$ is not the
pull-back of any matter-space two-form.

This second requirement selects the relevant decomposition, and it is here
that a purely metric or algebraic splitting is insufficient.  For a generic
non-null field one may always choose an orthonormal coframe
$\{\theta^\mu\}$ adapted to the eigenstructure of $F^a{}_b$, in which the
field takes the canonical algebraic form~\cite{Stazi2006}
\begin{equation}
F=\lambda\,\theta^0\wedge\theta^1+\mu\,\theta^2\wedge\theta^3 .
\label{eq:Fcanonical}
\end{equation}
Each term is simple, and the two principal two-planes
$\mathrm{span}\{\theta^0,\theta^1\}$ and
$\mathrm{span}\{\theta^2,\theta^3\}$ are intrinsic to $F_{ab}$: they are the
invariant eigenspaces of the antisymmetric endomorphism
$F^a{}_b=g^{ac}F_{cb}$~\cite{PenroseRindler1984}.  In this sense the
principal-plane splitting is metrically canonical.  However, the individual
summands in~\eqref{eq:Fcanonical} are in general \emph{not closed},
\begin{equation}
d\!\left(\lambda\,\theta^0\wedge\theta^1\right)\neq 0,
\qquad
d\!\left(\mu\,\theta^2\wedge\theta^3\right)\neq 0 ,
\end{equation}
even though their sum is closed by the Bianchi identity, $dF=0$.  The
coefficients $\lambda,\mu$ and the coframe $\theta^\mu$ vary over spacetime,
and the closedness of the total field only relates the two exterior
derivatives,
$d(\lambda\,\theta^0\wedge\theta^1)=-d(\mu\,\theta^2\wedge\theta^3)$,
without forcing either to vanish.  Consequently, the principal-plane
summands fail condition~\eqref{eq:sectorclosed} and cannot in general be the
pull-backs of matter-space two-forms.  The metric eigenplane decomposition
is therefore the wrong decomposition for the matter-space construction, even
though it is the natural one for displaying the Lorentz invariants.

The correct decomposition is supplied not by the metric but by the closed
structure of $F$ itself, through Darboux's theorem~\cite{Darboux}.  Since
$F=dA$ is closed and, in the generic non-null case, of constant rank four,
there exist local coordinates $(p_1,q_1,p_2,q_2)$ such that
\begin{equation}
F=dp_1\wedge dq_1+dp_2\wedge dq_2 .
\label{eq:Fdarboux}
\end{equation}
We then define the two sectors as the Darboux pairs
\begin{equation}
G^{(1)}=dp_1\wedge dq_1 ,
\qquad
G^{(2)}=dp_2\wedge dq_2 .
\label{eq:Gdarboux}
\end{equation}
Each sector is manifestly simple, $G^{(I)}\wedge G^{(I)}=0$, and at the same
time exact,
\begin{equation}
G^{(1)}=dX^{(1)},\quad X^{(1)}=p_1\,dq_1 ,
\qquad
G^{(2)}=dX^{(2)},\quad X^{(2)}=p_2\,dq_2 ,
\label{eq:Xdarboux}
\end{equation}
so that condition~\eqref{eq:sectorclosed} holds identically.  The Darboux
decomposition therefore yields precisely the closed simple sectors required
by the matter-space framework, with the sector potentials $X^{(I)}$ provided
automatically.  The mixed wedge product is
\begin{equation}
F\wedge F=2\,G^{(1)}\wedge G^{(2)}
=2\,dp_1\wedge dq_1\wedge dp_2\wedge dq_2 ,
\end{equation}
which is nonzero precisely when $F$ has rank four.

The price of using Darboux coordinates rather than the metric eigenframe is
that the decomposition~\eqref{eq:Gdarboux} is no longer metrically
canonical.  Darboux coordinates are far from unique: any symplectomorphism
preserving the form~\eqref{eq:Fdarboux} produces an equally valid pair of
closed simple sectors.  This non-uniqueness is, however, exactly the freedom
already present in the matter-space description, where the sector potentials
$X^{(I)}$ are defined only up to the restricted matter-space gauge
transformations of Sec.~\ref{subsec:matterspacegauge}.  What is invariant
under this freedom is the inter-sector four-form
$F\wedge F=2\,G^{(1)}\wedge G^{(2)}$, and it is this quantity, rather than any
particular choice of Darboux pair, that carries the physical content of the
decomposition.

It is worth noting that the Darboux decomposition is adapted to exactly the
invariant that is central to this paper.  The four-form $F\wedge F$ is
metric-independent: in the Darboux coordinates~\eqref{eq:Fdarboux} it is
simply twice the symplectic volume
$dp_1\wedge dq_1\wedge dp_2\wedge dq_2$, and it coincides with the
pseudoscalar invariant $F_{ab}{}^{*}F^{ab}$ up to the volume form.  By
contrast, the scalar invariant $F_{ab}F^{ab}$ involves the metric explicitly
and is not simply read off in Darboux coordinates.  The two decompositions
are thus adapted to different invariants: the metric
eigenframe~\eqref{eq:Fcanonical} organizes the scalar and pseudoscalar
invariants through $\lambda$ and $\mu$,
\[
F_{ab}F^{ab}\sim\mu^2-\lambda^2,
\qquad
F_{ab}{}^{*}F^{ab}\sim\lambda\mu ,
\]
whereas the Darboux decomposition organizes the closed simple structure and
the mutual invariant $F\wedge F$.  Only the latter is compatible with the
matter-space realization.

It is useful to emphasize that this two-flow decomposition is distinct from
both the observer-dependent electric--magnetic split and the complexified
self-dual/anti-self-dual split.  The electric--magnetic split,
\[
E_a=F_{ab}u^b,
\qquad
B_a={}^*F_{ab}u^b ,
\]
is a $3+1$ decomposition of a single spacetime two-form relative to an
observer $u^a$; by contrast, each Darboux sector $G^{(I)}$ is itself a
spacetime two-form and generically carries both electric and magnetic parts
with respect to a given observer.  The self-dual/anti-self-dual split,
\[
F=F^+ + F^-,
\qquad
F^\pm=\frac12\left(F\mp i*F\right),
\qquad
*F^\pm=\pm iF^\pm
\]
in Lorentzian signature, keeps the representation theory of two-forms
manifest but is complex.  The Darboux decomposition instead keeps the field
real and, crucially, closed sector by sector, which is the property demanded
by the matter-space construction.

Finally, the Darboux decomposition is local and assumes constant rank.  It
holds on an open region where $F$ is generic and non-null.  At loci where
the field becomes null or degenerate, or where the rank of $F$ changes, the
Darboux normal form~\eqref{eq:Fdarboux} changes type, and the two-flow
presentation must be patched across such loci or replaced by a null-adapted
one-flow description~\cite{Robinson1961,Trautman1977,Ranada1989}.  The
obstruction is therefore not the local existence of a closed simple
decomposition, which Darboux's theorem guarantees, but the global
continuation of a chosen Darboux sector basis through regions where the
algebraic type of $F_{ab}$ changes.  The global classification of these
transition loci is a separate problem and is not pursued here.

%==========================================
\section{From closed simple sectors to matter-space sectors}
\label{sec:principalplanesmatterspace}

A single matter-space sector consists of a map from spacetime to a three-dimensional label space \(M^A\) and a set of intrinsic matter-space fields which are pulled back to spacetime along the flow.
%For a single sector we denote the matter-space coordinates by \(M^A\).
The electromagnetic degrees of freedom are encoded in matter-space forms such as a one-form
\[
X=X_A(M) dM^A ,
\]
and a two-form
\[
G=\frac12 G_{AB}(M) dM^A\wedge dM^B .
\]

In the one-flow construction~\cite{Ho2026CQG}, the relations obeyed by the
intrinsic matter-space forms $X$ and $G$ are not imposed by hand; they are
derived from the structure of that theory.  First, once the matter-space
scalar $\phi$ is identified as a gauge degree of freedom, the absence of
magnetic monopoles together with the electromagnetic duality symmetry
implies that the intrinsic two-form is closed,
\be{closed}
dG=0 .
\ee
Second, the identities that follow from the dimensional difference between
the three-dimensional matter space and the four-dimensional spacetime then
yield the potential representation
\be{G:dX}
G=dX ,
\ee
where $X$ is the intrinsic matter-space one-form of the sector and $d$ is
the exterior derivative on the matter space.  We do not rederive
\eqref{closed} and \eqref{G:dX} here, but take them as part of the
matter-space structure inherited from the one-flow construction.

The gauge freedom is generated intrinsically by
\[
X\rightarrow X+d\phi ,
\]
where $\phi$ is the matter-space scalar, the 0-form.
The matter-space volume form,
\[
M=\frac1{3!}M_{ABC}(M)dM^A\wedge dM^B\wedge dM^C ,
\]
defines the associated conserved matter-space current $m^a$ after pull-back and Hodge dualization.  
Thus the basic one-flow matter-space data may be summarized as
\[
\bigl(M^A;\ \phi,\ X_A,\ G_{AB},\ M_{ABC}\bigr).
\]

The two-flow construction used below consists of two independent copies of this matter-space structure,
\[
\bigl(M_{(I)}^A;\ \phi^{(I)},\ X_A^{(I)},\ G_{AB}^{(I)},
M_{ABC}^{(I)}\bigr),
\qquad I=1,2 .
\]
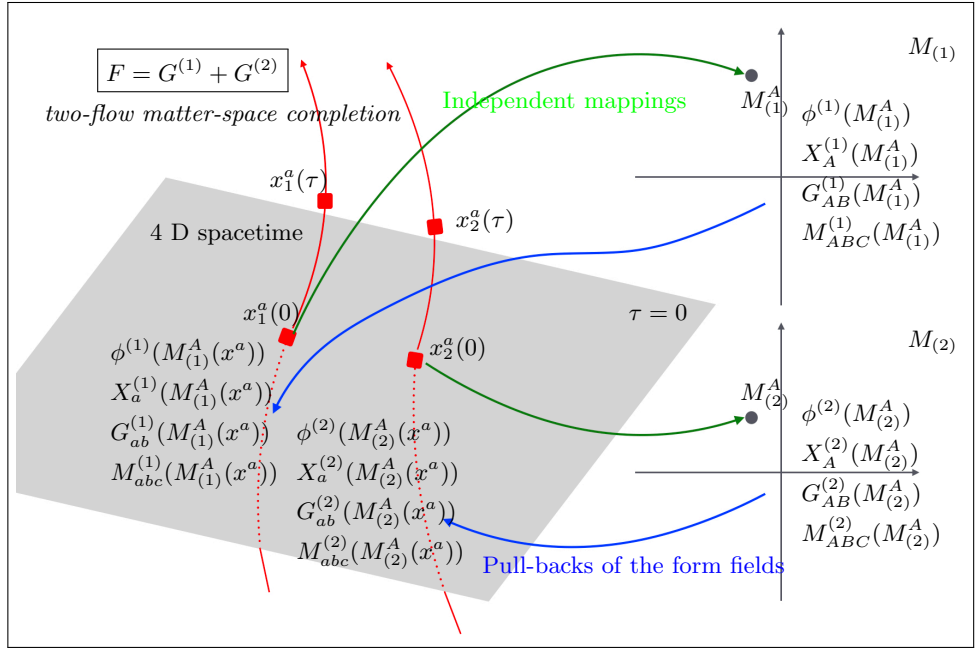
\begin{figure}[hbt]
\centering
 \boxed{\includegraphics[width=0.7\textwidth]{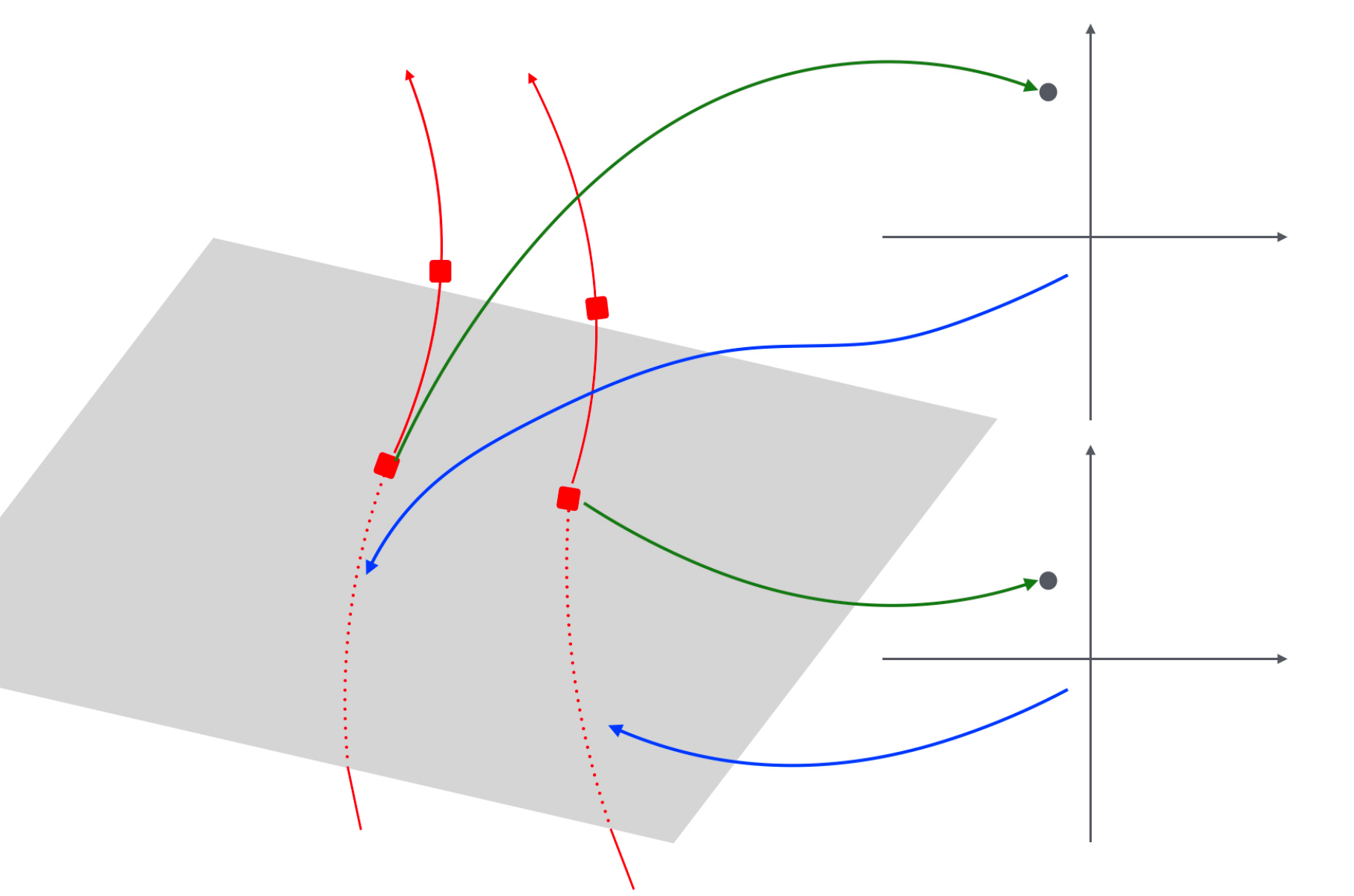}}
\put (-185,25) {\tcb{Pull-backs of the form fields}} 
\put (-200,200) {\tcg{Independent mappings}} 
\put (-310,150) {$4$ D spacetime } 
\put (-130, 120) {$\tau=0$}
\put (-275,120) {$x_1^a(0)$}
\put (-265,170) {$x_1^a(\tau)$}
\put (-25,220) {$M_{(1)}$ } 
\put (-88,200) {$M_{(1)}^A$ } 
\put (-65,195) {$\phi^{(1)}(M_{(1)}^A)$}
\put (-65,180) {$X_A^{(1)}(M_{(1)}^A)$}
\put (-65,165) {$G_{AB}^{(1)}(M_{(1)}^A)$}
\put (-65,150) {$M_{ABC}^{(1)}(M_{(1)}^A)$}
\put (-325,105) {$\phi^{(1)}(M_{(1)}^A(x^a))$}
\put (-325,90) {$X_a^{(1)}(M_{(1)}^A(x^a))$}
\put (-325,75) {$G_{ab}^{(1)}(M_{(1)}^A(x^a))$}
\put (-325,60) {$M_{abc}^{(1)}(M_{(1)}^A(x^a))$}
\put (-205,107) {$x_2^a(0)$}
\put (-195,155) {$x_2^a(\tau)$}
\put (-25,110) {$M_{(2)}$ }
\put (-88,90) {$M_{(2)}^A$ }  
\put (-65,82) {$\phi^{(2)}(M_{(2)}^A)$}
\put (-65,67) {$X_A^{(2)}(M_{(2)}^A)$}
\put (-65,52) {$G_{AB}^{(2)}(M_{(2)}^A)$}
\put (-65,37) {$M_{ABC}^{(2)}(M_{(2)}^A)$}
\put (-255,75) {$\phi^{(2)}(M_{(2)}^A(x^a))$}
\put (-255,60) {$X_a^{(2)}(M_{(2)}^A(x^a))$}
\put (-255,45) {$G_{ab}^{(2)}(M_{(2)}^A(x^a))$}
\put (-255,30) {$M_{abc}^{(2)}(M_{(2)}^A(x^a))$}
\put (-330,210) {$\boxed{F = G^{(1)} +G^{(2)}}$}
\put (-350,195) {\it two-flow matter-space completion}
\caption{
 Schematic structure of the two-flow matter-space completion.
Each of two independent three-dimensional matter spaces (right) is mapped into spacetime (left) by its own flow $M_{(I)}^A$ (green), and the intrinsic sector data $(\phi^{(I)},X_A^{(I)},G_{AB}^{(I)},M_{ABC}^{(I)})$ are pulled back along these maps (blue). A single sector yields a simple pull-back two-form, $G^{(I)}\wedge G^{(I)}=0$; the total field $F=G^{(1)}+G^{(2)}$ recovers the nondegenerate invariant only through the mixed pairing $F\wedge F=2G^{(1)}\wedge G^{(2)}$.
}
\label{fig:twoflow_schematic}
\end{figure}
The schematic plot is given in Fig.~\ref{fig:twoflow_schematic}.
Each copy has its own flow, its own matter-space gauge freedom, and its own intrinsic two-form.  
The spacetime two-forms \(G^{(I)}_{ab}\) are obtained by pulling back \(G^{(I)}_{AB}\) with the corresponding map
\(M_I^A(x)\).  
The total electromagnetic field is then not assigned to a single matter space, but is reconstructed as in Eq.~\eqref{eq:Fsum}.
Because the sector fields are closed by construction~\eqref{closed}, the total automatically satisfies the Bianchi identity,
\[
dF=0 .
\]
We do not consider decompositions of a closed total field into summands satisfying \(dG^{(1)}=-dG^{(2)}\neq0\) because they do not belong to one-flow sector.

For later convenience, we explicitly prescribe the definition of the matter-space current $m_{(I)}^a$
\begin{equation} \label{ma}
m^a_{(I)} = \frac1{3!} \epsilon^{abcd} M_{bcd} ^{(I)}
	= \frac1{3!} \epsilon^{abcd} (\nabla_bM^B_{(I)})
	(\nabla_cM^C_{(I)})(\nabla_dM^D_{(I)}) M^{(I)}_{BCD}.
\end{equation}
Here \(M^{(I)}_{abc}\) is the spacetime pull-back of the
matter-space volume form. Since \(M^{(I)}\) is a pulled-back three-form from a three-dimensional matter space, it is closed, and therefore conserved,
\be{m:conservation}
\nabla_a m^a_{(I)}=0 .
\ee
Moreover, \(m^a_{(I)}\) is tangent to the curves on which the matter
labels are constant,
\[
m^a_{(I)}\nabla_a M_{(I)}^A=0 .
\]
Consequently, any form pulled back from the \(I\)-th matter space is horizontal with respect to \(m^a_{(I)}\). In particular,
\be{eq:mGsector}
 m^a_{(I)}X^{(I)}_a=0, \qquad m^a_{(I)}G^{(I)}_{ab}=0 .
 \ee
In the one-flow system this current is naturally identified with the
helicity current~\cite{Ho2026CQG}. In the present two-flow construction, the two
currents \(m^a_{(I)}\) will accordingly be interpreted as the
self-helicity currents of the two matter-space sectors in
Sec.~\ref{sec:selfmutualGnotation}. 
Finally, since the sector gauge parameter is a pulled-back matter-space scalar,
\be{mdphi}
m^a_{(I)}\nabla_a\phi^{(I)}=0 ,
\ee
which characterizes the restricted matter-space gauge symmetry discussed below in Sec.~\ref{subsec:matterspacegauge}.

The decomposition described in Sec.~\ref{sec:algdecomposition} is a spacetime statement.
For a generic rank-four electromagnetic two-form, Darboux's theorem provides
local coordinates in which \(F=dp_1\wedge dq_1+dp_2\wedge dq_2\), identifying
two simple \emph{and closed} spacetime sectors as in Eq.~\eqref{eq:Gdarboux}.
Each sector is simple and hence satisfies
\begin{equation}
G^{(I)}\wedge G^{(I)}=0 ,
\end{equation}
and is closed,
\begin{equation}
dG^{(I)}=0 ,
\end{equation}
as required for a matter-space realization.

The role of the matter-space construction is to provide a pull-back
realization of these two simple spacetime sectors.  More explicitly, the
two-flow interpretation requires that each \(G^{(I)}\) can be written as
the spacetime projection of a two-form on an independent
three-dimensional matter space:
\begin{equation}
G^{(I)}_{ab}
=
(\nabla_a M_{(I)}^A)(\nabla_b M_{(I)}^B)
G^{(I)}_{AB},
\qquad
I=1,2 .
\label{eq:GIpullbackbridge}
\end{equation}
Equivalently, at the level of differential forms,
\begin{equation}
G^{(I)}
=
\frac12
G^{(I)}_{AB}(M_{(I)})
\,dM_{(I)}^A\wedge dM_{(I)}^B ,
\label{eq:GIpullbackforms}
\end{equation}
where the differentials \(dM_{(I)}^A\) are understood as spacetime
one-forms \(dM_{(I)}^A=(\nabla_a M_{(I)}^A)dx^a\).

Thus the Darboux decomposition and the matter-space
description refer to two different levels of the construction.  The
Darboux coordinates \((p_i,q_i)\) identify the closed simple sectors
selected by \(F_{ab}\), while the matter-space coordinates \(M_{(I)}^A\)
provide a material realization of each simple sector.  In this sense,
the Darboux decomposition~\eqref{eq:Fdarboux} is not replaced by the matter-space
description; rather, it is lifted to
\begin{equation}
dp_1\wedge dq_1
=
\frac12
G^{(1)}_{AB}(M_{(1)})
\,dM_{(1)}^A\wedge dM_{(1)}^B , 
\qquad
dp_2\wedge dq_2
=
\frac12
G^{(2)}_{AB}(M_{(2)})
\,dM_{(2)}^A\wedge dM_{(2)}^B .
\end{equation}
The two closed simple sectors are therefore interpreted as the spacetime
images of two independent matter-space pull-back sectors.

Since the matter space is three-dimensional, every matter-space
two-form \(G^{(I)}\) is decomposable on the open set where it is
nonzero.  Hence its spacetime pull-back is a simple two-form.  The
sector-adapted choice amounts to requiring this simple
pull-back to coincide with one of the two closed simple summands of
the spacetime field:
\[
(G^{(1)})_{\rm pb}=dp_1\wedge dq_1,
\qquad
(G^{(2)})_{\rm pb}=dp_2\wedge dq_2.
\]
This is a condition on the pulled-back two-forms, not on the individual
matter-space coordinate one-forms \(dM_{(I)}^A\).

%=====================================================
\subsection{Matter-space obstruction and the need for two flows}
\label{subsec:matterspaceobstruction}
%=====================================================
\emph{A single matter-space sector cannot represent a generic electromagnetic field}.  
To see this, consider a matter-space two-form
\[
G=\frac12 G_{AB}dM^A\wedge dM^B
\]
and its spacetime pull-back \(G_{ab}\).  In the preferred frame
\(F_{ab}=G_{ab}\).  Since the matter space is three-dimensional, the
four-form \(G\wedge G\) vanishes identically.  Hence
\[
F\wedge F
=
G_{\rm pb}\wedge G_{\rm pb}
=
(G\wedge G)_{\rm pb}
=
0 .
\]
Therefore every one-flow field lies in the degenerate sector
\(F\wedge F=0\).  Conversely, a field with \(F\wedge F\neq 0\) cannot be
the pull-back of a single matter-space two-form.

This result gives the precise sense in which the one-flow construction
is restricted.  The obstruction to a single matter-space realization is
not dynamical but kinematical: it follows from the absence of four-forms
on a three-dimensional matter space.  Therefore a generic electromagnetic
field requires at least two independent matter-space pull-back sectors,
as in \eqref{eq:Fdarboux}.

Matter space now plays a stronger role than merely motivating an
algebraic decomposition.  A decomposition~\eqref{eq:Fsum} is admissible in the present framework only if each summand can be realized as a matter-space pull-back~\eqref{eq:GIpullbackbridge}.
Thus the elementary sectors are not arbitrary simple bivectors; they are
pull-back two-forms carried by independent matter spaces.

%=====================================================
\subsection{Matter-space gauge sector}
\label{subsec:matterspacegauge}
%=====================================================

For each matter space the intrinsic one-form $ X^{(I)} $
is pulled back to spacetime as
\begin{equation}
X^{(I)}_a
=
(\nabla_a M_{(I)}^A)X^{(I)}_A .
\end{equation}
As noted above, the pulled-back sector forms are horizontal with respect to their own matter-space currents as shown in Eq.~\eqref{eq:mGsector}.

The matter-space gauge transformation is generated by a matter-space
scalar \(\phi^{(I)}(M_{(I)}^A)\):
\begin{equation}
X^{(I)}
\mapsto
X^{(I)}+d\phi^{(I)} .
\end{equation}
On spacetime this becomes
\begin{equation} \label{X-transform}
X^{(I)}_a
\mapsto
X^{(I)}_a+\nabla_a\phi^{(I)} ,
\end{equation}
where
\begin{equation}
\phi^{(I)}(x^a)=\phi^{(I)}(M_{(I)}^A(x^a)).
\end{equation}
Since \(\phi^{(I)}\) is pulled back from the \(I\)-th matter space, it obeys Eq.~\eqref{mdphi}.
Thus the sector gauge transformations are not arbitrary spacetime gauge
transformations.  They are precisely the gauge transformations compatible
with the corresponding matter-space flow.

The pull-back of Eq.~\eqref{G:dX} presents in spacetime
\begin{equation}
G^{(I)}_{ab}
=
\nabla_a X^{(I)}_b-\nabla_b X^{(I)}_a .
\label{eq:sectorFfromX}
\end{equation}
Thus, sector by sector, the matter-space one-form \(X^{(I)}_a\) plays the
role of a sector gauge potential, up to the restricted matter-space gauge
freedom.
The usual electromagnetic gauge potential is then identified with the
sum of the two sector potentials:
\begin{equation}
A_a
:=
X^{(1)}_a+X^{(2)}_a .
\label{eq:AtotalfromX}
\end{equation}
The total field strength becomes
\begin{align}
F_{ab}
&=
G^{(1)}_{ab}+G^{(2)}_{ab}
\nonumber\\
&=
\nabla_a X^{(1)}_b-\nabla_b X^{(1)}_a
+
\nabla_a X^{(2)}_b-\nabla_b X^{(2)}_a
\nonumber\\
&=
\nabla_a A_b-\nabla_b A_a .
\label{eq:FfromAtotal}
\end{align}
Therefore \(A_a\) has precisely the role of the conventional
electromagnetic gauge potential for the total field \(F_{ab}\).

Under independent matter-space gauge transformations~\eqref{X-transform} in the two sectors, the total potential transforms as
\begin{equation}
A_a
\mapsto
A_a+\nabla_a\Phi ,
\qquad
\Phi:=\phi^{(1)}+\phi^{(2)} .
\label{eq:Atotalgauge}
\end{equation}
This has the form of the usual electromagnetic gauge transformation of
the total potential.  However, \(\Phi\) should not be interpreted as a
matter-space scalar associated with either individual flow. 
Each \(\phi_I\) is adapted only to
its own matter-space direction \(m^a_{(I)}\), and the sum \(\Phi\) need
not satisfy
\[
m^a_{(I)}\nabla_a\Phi=0
\]
for either sector separately.  This does not represent a failure of the
gauge interpretation.  Rather, it reflects the fact that the generic
two-flow field has no single preferred matter-space direction.  Indeed,
for a nondegenerate total field,
$
F_{ab}=G^{(1)}_{ab}+G^{(2)}_{ab},
$
there is in general no nonzero vector \(n^a\) satisfying
$$
n^aF_{ab}=0 .
$$
Thus no vector \(n^a\) can play the role of a one-flow direction for the total field.

Consequently, the total gauge scalar \(\Phi\) should be regarded as an ordinary spacetime gauge parameter induced by the two sector gauge freedoms, not as a gauge scalar living on a single matter space.  
The sector potentials retain their restricted matter-space gauge symmetry, whereas their sum behaves as the usual electromagnetic gauge potential.

In this sense, the two-flow completion restores the ordinary gauge
behavior of the total electromagnetic potential.  
The restriction \(m^a\nabla_a\phi=0\) is a one-flow condition tied to the existence of a degenerate direction of the field.  
Once the two sectors are combined into a generic nondegenerate field, there is no single matter-space flow direction to which the total gauge scalar must be adapted.  
The induced transformation~\eqref{eq:Atotalgauge} therefore reproduces the ordinary gauge transformation of electromagnetism at the level of the total field.

%=====================================================
\section{Self and mutual helicity in matter-space notation}
\label{sec:selfmutualGnotation}
%=====================================================

In the one-flow theory the preferred current \(m^a\) is naturally
identified, up to normalization, with the helicity current. 
The use of helicity as a measure of electromagnetic or magnetic-field topology is standard in plasma physics and magnetohydrodynamics~\cite{FinnAntonsen1985,Berger1999}.
In the two-flow theory each matter space carries its own helicity sector.  For
the \(I\)-th sector we define the helicity three-form
\begin{equation}
\mathcal H_I
=
X^{(I)}\wedge G^{(I)} ,
\label{eq:sectorhelicityG}
\end{equation}
whose spacetime pull-back is
\begin{equation}
(\mathcal H_I)_{abc}
=
3X^{(I)}_{[a}G^{(I)}_{bc]} .
\end{equation}
The associated self-helicity current~\cite{BliokhBekshaevNori2013,AfanasievStepanovsky1996}, 
\begin{equation}
H^a_{(I)}
=
\frac{1}{3!}\epsilon^{abcd}(\mathcal H_I)_{bcd},
\end{equation}
 is conserved:
\begin{equation}
\nabla_a H^a_{(I)}=0 .
\end{equation}
Indeed,
\[
d\mathcal H_I
=
dX^{(I)}\wedge G^{(I)}
-
X^{(I)}\wedge dG^{(I)}
=
G^{(I)}\wedge G^{(I)}
=
0 ,
\]
where the last equality follows from the simplicity of the one-flow
sector.
Moreover, \(H^a_{(I)}\) satisfies the same algebraic conditions as the matter-space current \(m^a_{(I)}\):
\begin{equation}
H^a_{(I)}G^{(I)}_{ab}=0,
\qquad
H^a_{(I)}X^{(I)}_a=0 .
\end{equation}
These are precisely the analogues of Eq.~\eqref{eq:mGsector}.
Thus, on a generic one-flow sector where these conditions determine a
single flow direction, the matter-space current and the helicity current
are parallel:
\begin{equation}
m^a_{(I)}\parallel H^a_{(I)} .
\end{equation}
Since both currents are conserved, their relative normalization is fixed
up to a constant choice of flux normalization.  
We therefore identify the matter-space current with the self-helicity current, writing
\begin{equation}
m^a_{(I)}=H^a_{(I)}
\end{equation}
after this constant normalization has been chosen.

The helicity three-form of the total two-flow field is
\begin{equation}
\mathcal H
=
X\wedge F ,
\qquad
X:=X^{(1)}+X^{(2)},
\qquad
F:=G^{(1)}+G^{(2)} .
\end{equation}
Expanding gives
\begin{align}
\mathcal H
&=
X^{(1)}\wedge G^{(1)}
+
X^{(2)}\wedge G^{(2)}
+
X^{(1)}\wedge G^{(2)}
+
X^{(2)}\wedge G^{(1)} .
\end{align}
The first/last two terms are the self/mutual-helicity contributions of the two matter-space sectors in analogy with the self and mutual decomposition of magnetic helicity~\cite{BergerField1984,Candelaresi2021,SchuckLinton2023}.  
The last two terms define the mutual helicity three-form:
\begin{equation}
\mathcal H_{\rm mut}
=
X^{(1)}\wedge G^{(2)}
+
X^{(2)}\wedge G^{(1)} .
\label{eq:HmutG}
\end{equation}
This mutual helicity is not a form intrinsic to either matter space
separately.  It is defined only after both matter-space structures have been pulled back to the same spacetime.  

Taking the exterior derivative of \(\mathcal H_{\rm mut}\), one obtains
\begin{align}
d\mathcal H_{\rm mut}
&=
dX^{(1)}\wedge G^{(2)}
+
dX^{(2)}\wedge G^{(1)}
\nonumber\\
&=
2G^{(1)}\wedge G^{(2)} . \label{dHmut}
\end{align}
The wedge square of the total field is
\begin{align}
F\wedge F
&=
\left(G^{(1)}+G^{(2)}\right)
\wedge
\left(G^{(1)}+G^{(2)}\right)
=
2G^{(1)}\wedge G^{(2)} .  \label{FwedgeFmutual}
\end{align}
Using \eqref{FwedgeFmutual} with \eqref{dHmut}, we have
\begin{equation}
d\mathcal H_{\rm mut}=F\wedge F .
\label{eq:dHmutFG}
\end{equation}
Thus \(F\wedge F\) is not the mutual helicity itself. It is the
gauge-invariant four-form obtained as the exterior derivative of the
potential-level mutual helicity three-form.

%=====================================================
\subsection{ Nondegenerate fields require mutual helicity}
\label{subsec:mutualfreeGnotation}
%=====================================================
A natural question is whether the mutual helicity three-form itself can
be made to vanish:
\[
\mathcal H_{\rm mut}
=
X^{(1)}\wedge G^{(2)}
+
X^{(2)}\wedge G^{(1)}
=0 .
\]
From \(G^{(I)}=dX^{(I)}\), one has
\[
d\mathcal H_{\rm mut}=F\wedge F .
\]
Therefore a mutually helicity-free configuration,
$
\mathcal H_{\rm mut}=0
$
on an open region, implies
\[
F\wedge F=d\mathcal H_{\rm mut}=0 .
\]
Thus a two-flow configuration with vanishing mutual helicity cannot
represent a nondegenerate Maxwell field on that region. Equivalently,
when \(F\wedge F\neq0\), no choice of sector potentials can make
\(\mathcal H_{\rm mut}\) vanish identically. Mutual helicity is therefore
required in the two-flow realization of the generic
rank-four sector.

%==========================================
\subsection{Energy, helicity, and linking of the two sectors}
\label{subsec:mutualhelicityFwedgeF}

Equation~\eqref{eq:dHmutFG} identifies the invariant four-form
\(F\wedge F\) as the exterior derivative of the mutual helicity
three-form.
Thus the mutual helicity itself is a potential-level quantity, represented by the three-form~\eqref{eq:HmutG}, whereas \(F\wedge F\) is the gauge-invariant local source associated with it.  In this precise sense, \(F\wedge F\)  measures the
failure of \(\mathcal H_{\rm mut}\) to be closed.

For any spacetime region \(V\), Stokes' theorem gives
\begin{equation}
\int_V F\wedge F
=
\int_V d\mathcal H_{\rm mut}
=
\int_{\partial V}\mathcal H_{\rm mut}.
\end{equation}
The integral of \(F\wedge F\) over a four-volume therefore measures the
net mutual helicity flux through the boundary of that region.  It is a
creation-rate density for mutual helicity, not a helicity charge itself.

This interpretation is reminiscent of the role of the same pseudoscalar density in the chiral anomaly~\cite{Adler1969,BellJackiw1969}. 
For a massless charged fermion, the axial current is not conserved in an electromagnetic background, but obeys
schematically
\begin{equation}
\nabla_\mu j^\mu_5
\propto
F_{\mu\nu}{}^*F^{\mu\nu}.
\end{equation}
In differential-form language the same invariant is represented by the
four-form \(F\wedge F\).  The two statements should not be conflated:
the chiral anomaly concerns the nonconservation of a quantum axial
current, whereas \(\mathcal H_{\rm mut}\) is a classical potential-level
mutual helicity three-form.  Nevertheless, both are controlled by the
same invariant four-form.

In an observer decomposition, this four-form is proportional, up to
orientation and convention-dependent factors, to
\begin{equation}
(\mathbf E\cdot\mathbf B)\,d^4x .
\end{equation}
Thus \(\mathbf E\cdot\mathbf B\) is the observer-language expression of
the gauge-invariant mutual-helicity source.  In the present two-flow
formulation it should be interpreted as an inter-sector pseudoscalar
overlap, not as a self-helicity density of either individual sector.

This observation sharpens the physical meaning of the multi-flow
completion.  Each individual matter-space sector is simple:
\begin{equation}
G^{(I)}\wedge G^{(I)}=0 .
\end{equation}
Consequently, a single one-flow sector cannot generate a nonzero
\(F\wedge F\).  A generic rank-four Maxwell field requires a genuinely
two-flow configuration with a nonvanishing inter-sector pairing,
\begin{equation}
G^{(1)}\wedge G^{(2)}\neq 0 .
\end{equation}
Since
$
F=G^{(1)}+G^{(2)},
$
one has
$
F\wedge F
=
2G^{(1)}\wedge G^{(2)} .
$
Therefore the invariant \(F\wedge F\) is entirely a mixed quantity in
this decomposition.

It is useful to distinguish two different notions of helicity.  A single
degenerate sector may carry a conserved self-helicity,
\begin{equation}
\mathcal H_I
=
X^{(I)}\wedge G^{(I)},
\qquad
d\mathcal H_I
=
G^{(I)}\wedge G^{(I)}
=
0 .
\end{equation}
The closedness of \(\mathcal H_I\) does not imply that the helicity
itself vanishes.  It may still encode the self-linking, twisting, or
knottedness of field lines within a single sector~\cite{Moffatt1969,BergerField1984}.  
This distinction is particularly important for knotted null fields, such
as electromagnetic Hopfions~\cite{Ranada1989,IrvineBouwmeester2008,Kedia2013}.  
These configurations satisfy \(F\wedge F=0\), and therefore belong
naturally to the degenerate, one-flow class in the present
classification.  Nevertheless, they may carry nonzero electromagnetic
helicity and exhibit nontrivial field-line linking~\cite{Ranada1989,Irvine2011}.  
Their linking is a self-helicity effect within a single degenerate
sector, not a mutual helicity effect between two matter-space sectors.

By contrast, the mutual helicity~\eqref{eq:HmutG} measures the inter-sector linking between the two simple sectors.  Its
exterior derivative is $F\wedge F=2G^{(1)}\wedge G^{(2)} .$
Thus \(F\wedge F\) is not a measure of all electromagnetic linking.
Rather, it is the local source rate for mutual helicity between the two
matter-space sectors.  A null or degenerate field with \(F\wedge F=0\)
may still possess nonzero self-helicity, whereas a nonzero \(F\wedge F\)
signals the local production of mutual helicity in a genuinely two-flow
configuration.

The mutual helicity has a natural interpretation as a covariant
generalization of topological linking.  In the thin flux-tube limit, it
reduces to the standard linking interpretation of helicity
\cite{Moffatt1969,BergerField1984}.  In the familiar three-dimensional
magnetic case,
\begin{equation}
H_{\rm mag}
=
\int_\Sigma A\wedge F
\end{equation}
measures the linkage of magnetic flux tubes.  For two thin flux tubes
with fluxes \(\Phi_1\) and \(\Phi_2\), the mutual contribution reduces to
\begin{equation}
H_{\rm mut}
=
2\,{\rm Lk}(C_1,C_2)\,\Phi_1\Phi_2 ,
\end{equation}
where \({\rm Lk}(C_1,C_2)\) is the linking number of the two tube
centrelines.  Related topological electromagnetic configurations,
including Hopfion and knotted light solutions, provide concrete examples
in which electromagnetic helicity and linking acquire a direct field-line
interpretation~\cite{Ranada1989,Kedia2013,ArrayasTrueba2019}.

In the present two-flow formulation, the mutual helicity
three-form~\eqref{eq:HmutG} plays the analogous role for the two simple
electromagnetic sectors.  Integrating Eq.~\eqref{eq:dHmutFG}, one obtains
\begin{equation}
\int_M 2G^{(1)}\wedge G^{(2)}
=
\int_{\partial M}\mathcal H_{\rm mut}.
\end{equation}
This identity shows that the inter-sector pairing in the bulk is
measured on the boundary as mutual helicity, in direct analogy with the
relation between helicity and linking in three-dimensional magnetic
fields.

One should not, however, identify the mutual helicity with an integer
linking number in complete generality.  It becomes a genuine topological
linking invariant only under suitable boundary conditions, flux-tube
localization, and gauge choices.  In a generic smooth electromagnetic
field it is better understood as a covariant linking density, or as a
linking-like helicity functional.

\vspace{.2cm}

The two-flow decomposition also gives rise to two distinct inter-sector
bilinear pairings.  The first is the metric pairing
\begin{equation}
G^{(1)}\wedge *G^{(2)} ,
\end{equation}
which appears in the Maxwell action and contributes directly to the
stress-energy tensor.  This term represents the dynamical coupling energy
between the two sectors.  The Maxwell stress-energy tensor of the total
field,
\begin{equation}
T_{ab}
=
F_{ac}F_b{}^c
-
\frac14 g_{ab}F_{cd}F^{cd},
\end{equation}
decomposes as
\begin{equation}
T_{ab}
=
T^{(1)}_{ab}
+
T^{(2)}_{ab}
+
T^{(\mathrm{cross})}_{ab},
\end{equation}
where
\begin{equation}
T^{(I)}_{ab}
=
G^{(I)}_{ac}G^{(I)}_{b}{}^{c}
-
\frac14 g_{ab}G^{(I)}_{cd}G^{(I)cd},
\end{equation}
and
\begin{equation}
T^{(\mathrm{cross})}_{ab}
=
G^{(1)}_{ac}G^{(2)}_{b}{}^{c}
+
G^{(2)}_{ac}G^{(1)}_{b}{}^{c}
-
\frac12 g_{ab}G^{(1)}_{cd}G^{(2)cd}.
\end{equation}
Thus the energy and momentum of the two-flow electromagnetic field are
not, in general, the sum of two independent sector contributions.  The
cross term represents the metric inter-sector coupling generated by
\(G^{(1)}\wedge *G^{(2)}\).

This structure is analogous to entrainment in multi-fluid dynamics.  In
an entrained multi-fluid system, the momentum of one constituent is not
determined solely by its own current, but also depends on the motion of
the other constituents.  Similarly, in the present two-flow
electromagnetic construction, the stress-energy tensor contains an
irreducible cross term mixing the two matter-space sectors.  The pairing
\(G^{(1)}\wedge *G^{(2)}\) is therefore the metric, or dynamical,
entrainment-like coupling between the sectors.

The second inter-sector pairing is the pseudoscalar pairing
\begin{equation}
G^{(1)}\wedge G^{(2)} .
\end{equation}
Unlike the metric pairing, it is independent of the Hodge dual and does
not represent the dynamical coupling energy.  Instead, it is the source
of mutual helicity:
\begin{equation}
G^{(1)}\wedge G^{(2)}
=
\frac12 d\mathcal H_{\rm mut}.
\end{equation}
The two pairings therefore encode complementary aspects of the same
two-flow structure:
\begin{align}
\text{metric entrainment-like coupling}
&\sim
G^{(1)}\wedge *G^{(2)},\\
\text{mutual-helicity source coupling}
&\sim
G^{(1)}\wedge G^{(2)} .
\end{align}
The former is metric and dynamical, while the latter is pseudoscalar and
helicity-like.  In observer language, the former corresponds to the
parity-even inter-sector overlap, whereas the latter corresponds to the
parity-odd inter-sector overlap already encoded in \(F\wedge F\).
Thus the coupling energy and the mutual helicity source are not
identical, but are two different projections of the inter-sector
structure.

\paragraph*{Summary:}
The preceding discussion gives the physical interpretation of the
two-flow completion.  A one-flow configuration represents a single
simple, self-helicity-carrying sector and is therefore necessarily
degenerate.  A generic nondegenerate Maxwell field is obtained only when
two such sectors are embedded in spacetime with a nontrivial mutual
pairing.  Thus the passage from the one-flow theory to the full Maxwell
field is not merely a doubling of matter-space currents; it is the
appearance of an inter-sector structure, measured locally by
\(G^{(1)}\wedge G^{(2)}\) or equivalently by \(F\wedge F\), that allows
the total field to escape the degenerate one-flow sector.

%=====================================================
\subsection{Example: a constant field with \(\mathbf E\parallel \mathbf B\)}
\label{subsec:exampleEparallelB}
%=====================================================

It is useful to illustrate the two-flow construction with the simplest
nondegenerate electromagnetic configuration.  Consider Minkowski
spacetime with coordinates \((t,x,y,z)\), and take a constant electric
field and a constant magnetic field both pointing in the \(z\)-direction:
\begin{equation}
\mathbf E=E\,\hat z,
\qquad
\mathbf B=B\,\hat z .
\end{equation}
With a conventional choice of orientation, the corresponding field
strength may be written as
\begin{equation}
F
=
E\,dt\wedge dz
+
B\,dx\wedge dy .
\label{eq:EparallelBfield}
\end{equation}
This field is not of one-flow type when \(EB\neq 0\), since
\begin{equation}
F\wedge F
=
2EB\,dt\wedge dz\wedge dx\wedge dy
\neq 0 .
\label{eq:EparallelBFwedgeF}
\end{equation}
Equivalently, the pseudoscalar invariant \(\mathbf E\cdot\mathbf B\) is
nonzero.

The two-flow decomposition is immediate.  Define
\begin{equation}
G^{(1)}
=
E\,dt\wedge dz,
\qquad
G^{(2)}
=
B\,dx\wedge dy .
\label{eq:EparallelBsectors}
\end{equation}
Each sector is simple:
\begin{equation}
G^{(1)}\wedge G^{(1)}=0,
\qquad
G^{(2)}\wedge G^{(2)}=0 .
\end{equation}
Nevertheless, their mixed wedge product is nonzero:
\begin{equation}
2G^{(1)}\wedge G^{(2)}
=
2EB\,dt\wedge dz\wedge dx\wedge dy .
\end{equation}
Thus $
F\wedge F
=
2G^{(1)}\wedge G^{(2)} .$
This explicitly realizes the general statement that the invariant which vanishes in each one-flow sector arises entirely from the mutual configuration of the two sectors.

This constant field is also the simplest illustration of the Darboux
decomposition of Sec.~\ref{subsec:darboux}: the coordinates
\((t,z,x,y)\) are Darboux coordinates for \(F\), with the identifications
\((p_1,q_1)=(Et,z)\) and \((p_2,q_2)=(Bx,y)\), so that each sector in
\eqref{eq:EparallelBsectors} is automatically closed,
\(dG^{(I)}=0\).

One can also display the corresponding local matter-space realization.
For the first sector, choose matter-space coordinates
\begin{equation}
M_{(1)}^1=t,\qquad
M_{(1)}^2=z,\qquad
M_{(1)}^3=x,
\end{equation}
and take the intrinsic matter-space two-form
\begin{equation}
G^{(1)}
=
E\,dM_{(1)}^1\wedge dM_{(1)}^2 .
\end{equation}
Its spacetime pull-back is
\begin{equation}
G^{(1)}
=
E\,dt\wedge dz .
\end{equation}
For the second sector, choose
\begin{equation}
M_{(2)}^1=x,\qquad
M_{(2)}^2=y,\qquad
M_{(2)}^3=t,
\end{equation}
with
\begin{equation}
G^{(2)}
=
B\,dM_{(2)}^1\wedge dM_{(2)}^2 .
\end{equation}
The pull-back gives
\begin{equation}
G^{(2)}
=
B\,dx\wedge dy .
\end{equation}
Hence the total field \eqref{eq:EparallelBfield} is obtained as the sum
of two matter-space pull-back sectors:
\begin{equation}
F=G^{(1)}+G^{(2)} .
\end{equation}

This example should be understood as a local realization of the
two-flow decomposition.  The individual choices of \(M_I^A\) are not
unique, and no physical significance should be assigned to this
particular coordinate representation.  What is invariant is the fact that
the two simple sectors are individually degenerate while their mutual
wedge product is nonzero.

The mutual helicity interpretation is also transparent.  
One may choose local sector potentials
\begin{equation}
X^{(1)}
=
E\,t\,dz,
\qquad
X^{(2)}
=
B\,x\,dy ,
\end{equation}
so that
\begin{equation}
dX^{(1)}=G^{(1)},
\qquad
dX^{(2)}=G^{(2)} .
\end{equation}
The mutual helicity three-form is then
\begin{equation}
\mathcal H_{\rm mut}
=
X^{(1)}\wedge G^{(2)}
+
X^{(2)}\wedge G^{(1)} .
\end{equation}
Taking the exterior derivative gives
\begin{equation}
d\mathcal H_{\rm mut}
=
2G^{(1)}\wedge G^{(2)}
=
F\wedge F .
\end{equation}

This example shows explicitly why a single matter-space sector is
insufficient for a generic electromagnetic field.  Each sector in
\eqref{eq:EparallelBsectors} is a perfectly valid one-flow-type simple
two-form, but neither one can produce a nonzero \(F\wedge F\) by itself.
The nondegenerate Maxwell field is obtained only after the two sectors
are superposed, and its nonzero invariant \(\mathbf E\cdot\mathbf B\) is entirely an inter-sector effect.

%=====================================================
\section{Two-flow action and variational equations}
\label{sec:twoflowaction}
%=====================================================

We now ask whether the two-flow completion leads back to the usual
Maxwell equation once the dynamics is supplied by an action principle.
The answer is affirmative, provided the action is written in terms of
the total electromagnetic field rather than as a sum of two independent
one-flow actions.

For each sector, $ G^{(I)}_{ab}$ is the spacetime pull-back of the intrinsic matter-space two-form \(G^{(I)}_{AB}\) as in Eq.~\eqref{eq:GIpullbackbridge}.  
In the preferred frame the sector field strengths are identified with these pull-backs, and the total electromagnetic field is given in Eq.~\eqref{eq:Fsum}.
%In the homogeneous branch, where each matter-space two-form is closed, one may locally write as in Eq.~\eqref{eq:sectorFfromX} with $dG^{(I)}=0$.
The total potential is then given by Eq.~\eqref{eq:AtotalfromX}
so that $F_{ab}=\nabla_a A_b-\nabla_b A_a .$

The minimal action is taken to be the Maxwell action for the total field,
coupled to the electric current \(j^a\):
\begin{equation}
\Lambda
=
-\frac14 F_{ab}F^{ab}
-
j^a A_a .
\label{eq:twoflowLagrangian}
\end{equation}
Equivalently, using \eqref{eq:Fsum} and \eqref{eq:AtotalfromX},
\begin{align}
\Lambda
&=
-\frac14
\left(G^{(1)}_{ab}+G^{(2)}_{ab}\right)
\left(G^{(1)ab}+G^{(2)ab}\right)
-
j^a\left(X^{(1)}_a+X^{(2)}_a\right)
\nonumber\\
&=
-\frac14G^{(1)}_{ab}G^{(1)ab}
-\frac14G^{(2)}_{ab}G^{(2)ab}
-\frac12G^{(1)}_{ab}G^{(2)ab}
-
j^aX^{(1)}_a
-
j^aX^{(2)}_a .
\label{eq:twoflowLagrangianexpanded}
\end{align}
The cross term
\begin{equation}
-\frac12G^{(1)}_{ab}G^{(2)ab}
\end{equation}
is essential.  Without it the theory would describe two independent
degenerate one-flow sectors rather than a single electromagnetic field
formed from their superposition.

As in the one-flow construction~\cite{Ho2026CQG}, and in the standard
convective variational formulation of relativistic fluids~\cite{Carter1989,AnderssonComer2007,LK2022,Kim:2023lta}, we use a first-order variational description. 
The conjugate to the total field strength is
\begin{equation}
\Pi^{ab}
:=
\frac{\partial\Lambda}{\partial F_{ab}} .
\end{equation}
For the Maxwell Lagrangian \eqref{eq:twoflowLagrangian}, this gives
\begin{equation}
\Pi^{ab}
=
-\frac12F^{ab}
=
-\frac12\left(G^{(1)ab}+G^{(2)ab}\right).
\label{eq:twoflowPi}
\end{equation}

Since there are now two matter spaces, there are two independent
Lagrangian displacement vectors,
\begin{equation}
\xi^a_{(1)},
\qquad
\xi^a_{(2)}.
\end{equation}
For the \(I\)-th matter space, the Lagrangian variation satisfies
\begin{equation}
\Delta_I M_{(I)}^A=0,
\end{equation}
and hence the Eulerian variation is
\begin{equation}
\delta_I M_{(I)}^A
=
-\mathcal L_{\xi_{(I)}}M_{(I)}^A .
\end{equation}
Therefore the induced variations of the \(I\)-th sector fields are
\begin{equation}
\delta_I G^{(I)}_{ab}
=
-\mathcal L_{\xi_{(I)}}G^{(I)}_{ab},
\qquad
\delta_I X^{(I)}_a
=
-\mathcal L_{\xi_{(I)}}X^{(I)}_a .
\label{eq:sectorLievariation}
\end{equation}
The other sector is held fixed under this variation:
\begin{equation}
\delta_I G^{(J)}_{ab}=0,
\qquad
\delta_I X^{(J)}_a=0,
\qquad
I\neq J .
\end{equation}

Varying the action with respect to the \(I\)-th displacement
\(\xi^a_{(I)}\), and discarding total derivative terms, gives the
sector-wise Euler equation
\begin{equation}
X^{(I)}_e\nabla_a j^a
-
3\Pi^{ab}\nabla_{[e}G^{(I)}_{ab]}
+
2G^{(I)}_{ea}\nabla_b\Pi^{ba}
+
2j^a\nabla_{[a}X^{(I)}_{e]}
=
0 .
\label{eq:twoflowEulergeneral}
\end{equation}
If the electric current is conserved,
\begin{equation}
\nabla_a j^a=0,
\end{equation}
this reduces to
\begin{equation}
-
3\Pi^{ab}\nabla_{[e}G^{(I)}_{ab]}
+
2G^{(I)}_{ea}\nabla_b\Pi^{ba}
+
2j^a\nabla_{[a}X^{(I)}_{e]}
=
0 .
\label{eq:twoflowEulerconserved}
\end{equation}

Using Eq.~\eqref{eq:twoflowPi} and 
\begin{equation}
\nabla_{[e}G^{(I)}_{ab]}=0,
\qquad
2\nabla_{[a}X^{(I)}_{e]}=G^{(I)}_{ae},
\end{equation}
equation~\eqref{eq:twoflowEulerconserved} becomes
\begin{equation}
G^{(I)}_{ea}
\left(
\nabla_bF^{ab}-j^a
\right)
=
0,
\qquad
I=1,2 .
\label{eq:projectedMaxwellTwoFlow}
\end{equation}
Thus the variational principle does not give two independent Maxwell
equations.  It gives two projections of the same Maxwell equation, one
along each matter-space sector.

Let
\begin{equation}
\Xi^a
:=
\nabla_bF^{ab}-j^a .
\label{eq:Xidefinition}
\end{equation}
Then \eqref{eq:projectedMaxwellTwoFlow} states that
\begin{equation}
G^{(1)}_{ea}\Xi^a=0,
\qquad
G^{(2)}_{ea}\Xi^a=0 .
\end{equation}
Equivalently,
\begin{equation}
\Xi^a
\in
\ker G^{(1)}\cap \ker G^{(2)} .
\label{eq:kernelintersection}
\end{equation}
For a generic two-flow configuration, the two degenerate sectors have no
common kernel:
\begin{equation}
\ker G^{(1)}\cap\ker G^{(2)}=\{0\}.
\label{eq:trivialkernelintersection}
\end{equation}
In this case \eqref{eq:kernelintersection} implies
\begin{equation}
\Xi^a=0,
\end{equation}
and therefore the standard sourced Maxwell equation is recovered:
\begin{equation}
\nabla_bF^{ab}=j^a .
\label{eq:MaxwellRecoveredTwoFlow}
\end{equation}

This shows that the two-flow completion removes, in the generic case,
the kernel ambiguity present in the one-flow construction.  In the
one-flow theory the field strength is degenerate, and the variational
equation only fixes the component of \(\nabla_bF^{ab}-j^a\) transverse
to the kernel of \(F_{ab}\).  In the two-flow theory each sector is still
degenerate, but their kernels generically intersect trivially.  Hence
the two projected equations together force the full Maxwell equation.

There may nevertheless exist special two-flow configurations for which
\begin{equation}
\ker G^{(1)}\cap\ker G^{(2)}\neq \{0\}.
\end{equation}
In such cases the most general result of the variational equation is
\begin{equation}
\nabla_bF^{ab}
=
j^a+k^a,
\qquad
k^a\in \ker G^{(1)}\cap\ker G^{(2)} .
\label{eq:twoflowkernelcorrection}
\end{equation}
The residual current \(k^a\) is the two-flow analogue of the kernel
correction that appears in the one-flow theory.  It vanishes for generic
rank-four configurations but may survive in special degenerate cases.

The conclusion is that the correct two-flow action is not the sum of two
independent one-flow Maxwell actions.  It is the Maxwell action of the
total field
\[
F=G^{(1)}+G^{(2)}.
\]
With this choice, the inter-sector energy coupling is automatically
included, and the standard Maxwell equation follows from the two
matter-space variations whenever the combined field has no residual
common kernel.

%========================================\\
\section{Summary and Discussions}
\label{sec:discussion}
%========================================\\
In this work, we have shown that, locally in four spacetime dimensions, no more than two matter-space flows are required to represent Maxwell fields.  A single flow describes the degenerate sector \(F\wedge F=0\), while the second flow provides the minimal completion needed for the generic case.
The two-flow completion of the matter-space formulation may be summarized as follows.
\begin{center}
\renewcommand{\arraystretch}{1.35}
\begin{tabular}{c|c|c}
\hline\hline
 & One-flow matter space & Two-flow matter spaces \\
\hline
Field strength
&
\(F=G\)
&
\(F=G^{(1)}+G^{(2)}\)
\\

Matter spaces
&
one matter space \(M^A\)
&
two independent matter spaces \(M_{(1)}^A,M_{(2)}^A\)
\\

Degeneracy
&
\(F\wedge F=0\)
&
\(F\wedge F=2G^{(1)}\wedge G^{(2)}\)
\\

Elementary sectors
&
one simple two-form
&
two simple two-forms,
\(G^{(I)}\wedge G^{(I)}=0\)
\\

Gauge potential
&
\(A_a=X_a\)
&
\(A_a=X^{(1)}_a+X^{(2)}_a\)
\\

Gauge scalar
&
\(\phi(M^A)\),
\(m^a\nabla_a\phi=0\)
&
\(\Phi=\phi^{(1)}(M_{(1)}^A)+\phi^{(2)}(M_{(2)}^A)\)
\\

Helicity
&
self-helicity only
&
self-helicity \(+\) self-helicity \(+\) mutual helicity
\\

Lorentz invariant
&
\(F\wedge F=0\)
&
\(F\wedge F =2G^{(1)}\wedge G^{(2)}\) 
\\

Dynamics
&
\begin{tabular}{c}
	projected Maxwell equation \\
	with possible kernel ambiguity
\end{tabular}
&
\begin{tabular}{c}
generic recovery \\
of the full Maxwell equation
\end{tabular}
\\
\hline\hline
\end{tabular}
\end{center} 
In the two-flow case, the Lorentz invariant pseudoscalar, \(F\wedge F\), vanishes in each isolated sector, is generated entirely by the mixed wedge product between the two sectors.
A single matter-space sector carries an intrinsic two-form \(G_{AB}\), whose spacetime pull-back \(G_{ab}\) is identified with the electromagnetic field strength in the geometrically preferred frame. 
Because the matter space is three-dimensional, the corresponding four-form vanishes identically,
$
G\wedge G = 0,
$
and therefore the one-flow construction naturally selects the degenerate electromagnetic sector,
$
F\wedge F = 0.
$
In this sense, the one-flow theory does not describe a generic Maxwell field, but rather an elementary helicity-carrying sector represented by a simple two-form.

A generic electromagnetic field requires at least two independent matter-space sectors. 
Locally, the field strength may be written as
$
F = G^{(1)} + G^{(2)},
$
where each sector is individually simple and satisfies
$
G^{(I)}\wedge G^{(I)}=0.
$
As discussed in Sec.~\ref{subsec:darboux}, the admissible decomposition is the one supplied by Darboux's theorem, in which each summand is not only simple but closed, $dG^{(I)}=0$, and hence realizable as a matter-space pull-back. 
The full electromagnetic Lorentz invariant then arises entirely from the mixed term,
\begin{equation*}
F\wedge F = 2\,G^{(1)}\wedge G^{(2)},
\end{equation*}
absent from each isolated degenerate sector.
% but present in their coupled superposition.
Because each sector is locally exact,
the mixed term is naturally related to a mutual helicity three-form. 
Accordingly, the total helicity decomposes into two sector self-helicities together with an additional mutual contribution. 
We therefore conclude that the one-flow matter-space theory captures the elementary self-helicity sector of electromagnetism, while the full local Maxwell field is obtained through the minimal coupled completion by two such sectors. 
The invariant content of generic electromagnetism is then encoded in their mutual coupling.
The two-flow completion developed here provides a new geometric interpretation of the local structure of electromagnetism. 
Rather than viewing a generic Maxwell field as a single irreducible object, the present formulation resolves it into two individually degenerate matter-space sectors, each carrying its own helicity current, together with the mutual coupling required to produce a nondegenerate field. 
%From this viewpoint, the one-flow theory is not a failed attempt to describe all of electromagnetism with a single matter-space sector; rather, it identifies the minimal degenerate sector from which a generic field is assembled.

A particularly important consequence is that mutual helicity is not a
freely removable potential-level decoration.  
Setting the mutual contribution to zero forces \(F\wedge F=0\), so that the total field returns to the degenerate sector.  
Thus the nondegenerate structure of a generic Maxwell field is inseparable from the mutual spacetime embedding of two elementary helicity sectors.
This result is closely tied to the special algebraic structure of two-forms in four spacetime dimensions. 
In four dimensions, a generic non-null two-form is locally the sum of exactly two simple two-forms, and its nondegeneracy is already detected by the top-degree invariant \(F\wedge F\). 
Since each simple sector separately satisfies \(G^{(I)}\wedge G^{(I)}=0\), the full invariant structure must arise entirely from the inter-sector term. 
In this sense, mutual helicity is not merely a feature of a chosen
sector decomposition.  
It is the four-dimensional expression of the fact that a generic Maxwell field is the minimal coupled completion of two degenerate helicity sectors.

This observation also suggests a natural higher-dimensional perspective. 
In even spacetime dimension \(2n\), a generic two-form decomposes into \(n\) simple sectors, and the corresponding top-degree invariant is \(F^n\). 
From this viewpoint, the four-dimensional Maxwell field is the first nontrivial case in which genericity is controlled entirely by a pairwise coupling between two simple sectors. 
In higher dimensions, one should expect more general higher-order inter-sector couplings rather than a single mutual helicity term. 
The present two-flow construction therefore highlights not only the structure of electromagnetism in matter-space language, but also a distinctive feature of four-dimensional gauge theory itself.

The picture that emerges is therefore the following. 
A one-flow matter-space sector describes a single simple electromagnetic configuration and is naturally associated with a conserved self-helicity current. 
A generic electromagnetic field requires two such sectors, and its nontrivial invariant content is encoded in their mutual coupling. 
In this sense, four-dimensional electromagnetism is exceptional: the full local Maxwell field is obtained from the minimal coupling of just two elementary one-flow helicity sectors, and its genericity is measured precisely by their mutual interaction.

Compared with the standard formulation, the present construction offers three structural advantages.  
First, it gives a matter-space origin to the degeneracy condition \(F\wedge F=0\) of a one-flow sector.  
Second, it interprets the invariant \(F\wedge F\) of a generic field as an inter-sector quantity measuring the mutual embedding of two matter-space pull-back structures.  
Third, in the two-flow action principle, the standard sourced Maxwell equation is recovered generically because the intersection of the two sector kernels is trivial.

\paragraph*{Outlook: coupling to gravity and helicity dynamics.}
The two-flow structure acquires further physical significance once the
electromagnetic field is coupled to gravity. A detailed treatment lies
beyond the scope of this paper and is left to future work; here we only
indicate why the decomposition is expected to be useful in that setting.

In the present classification a single matter-space sector is degenerate,
$F\wedge F=0$, which in an observer frame is the condition $\mathbf E\cdot\mathbf B=0$. 
In covariant language this is the ideal-magnetohydrodynamical condition \(F_{ab}u^b=0\), which implies \(F\wedge F=0\), or equivalently \(\mathbf E\cdot\mathbf B=0\) in the comoving frame;
it also contains null electromagnetic radiation as a limiting case. In
this sector the source of helicity vanishes kinematically, $d\mathcal
H=F\wedge F=0$, so that the self-helicity is an exactly conserved
topological charge. Because the obstruction is purely algebraic, this
conservation persists in a curved and even dynamical spacetime, reducing
in a cosmological background to the conservation of the comoving magnetic
helicity. The one-flow sector is therefore the natural geometric home of
the ideal, helicity-conserving regime.

A single sector cannot, however, accommodate $\mathbf E\cdot\mathbf
B\neq0$. This is a physical rather than a technical limitation, since the
leading mechanisms that generate or transfer magnetic helicity in the
early universe operate precisely through $\mathbf E\cdot\mathbf B\neq0$:
inflationary magnetogenesis, including parity-violating couplings of
the form \(\phi F\wedge F\)~\cite{TurnerWidrow1988,GarretsonFieldCarroll1992,AnberSorbo2010}, and the chiral magnetic effect
and chiral anomaly, by which fermion chirality is converted into magnetic helicity \cite{JoyceShaposhnikov1997,BoyarskyFrohlichRuchayskiy2012,BoyarskyFrohlichRuchayskiy2015}. 
A one-flow electromagnetic field cannot describe the \(F\wedge F\neq0\) part of these processes.
%None of these can be represented by a one-flow field. 
The two-flow completion is the minimal framework that
opens this channel, since $F\wedge F=2\,G^{(1)}\wedge G^{(2)}=d\mathcal
H_{\rm mut}$ may be nonzero, while two separately conserved self-helicity
currents are retained. Helicity generation is then cleanly isolated as the
production of mutual helicity between two otherwise ideal sectors.
Furthermore, the two inter-sector pairings of Sec.~\ref{subsec:mutualhelicityFwedgeF} separate the
way the field sources gravity: 
the metric pairing \(G^{(1)}\wedge{}^{*}G^{(2)}\) is parity-even and
contributes to the stress-energy tensor, and hence to the metric
response, including the gravitational-wave sector when the corresponding anisotropic stresses are present;
%the metric pairing $G^{(1)}\wedge{}^{*}G^{(2)}$ is parity-even and feeds the stress-energy tensor, and hence the gravitational (e.g.\ gravitational-wave) response, 
whereas the pseudoscalar
pairing $G^{(1)}\wedge G^{(2)}$ is parity-odd and controls helicity
production. We therefore expect the two-flow formulation to provide a
useful bookkeeping for helical magnetogenesis and for parity-odd
gravitational signatures, in which the conserved (ideal) and the generated
(anomalous) parts of the helicity are carried by distinct, geometrically
defined variables. A concrete Einstein--Maxwell analysis along these
lines---including the backreaction of the cross stress
$T^{(\mathrm{cross})}_{ab}$ and the evolution of the two self-helicity
charges in a cosmological background---will be presented elsewhere.

%=================================\\
\begin{acknowledgments}
This work was supported by the National Research Foundation of Korea (NRF) grant with grant numbers RS-2026-25483539 (H.K.).
AI-assisted tools were used only for language editing; all analytic derivations and scientific conclusions are the author's responsibility.
\end{acknowledgments}

\appendix
%========

%=========================================
%=========================================

\end{document}